\documentclass[a4paper, authoryear]{article}
\usepackage{techRepsPDF}

\usepackage[english]{babel}
\usepackage{tikz}
\usetikzlibrary{arrows}
\usepackage{graphicx}
\usepackage{color}
\usepackage{mathptmx}      
\usepackage{amsmath}

\usepackage{siunitx}	

\usepackage{url}
\usepackage{array}

\usepackage[linesnumbered,ruled,vlined,]{algorithm2e}

\SetCommentSty{mycommfont}
\newcommand{\maxim}{\mbox{maximize }} 

\newcommand{\setH}{\mathcal{H}}
\newcommand{\setI}{\mathcal{I}}

\newcommand{\setT}{\mathcal{T}}
\newcommand{\setS}{\mathcal{S}}
\newcommand{\ie}{\emph{i.e.}}
\newcommand{\eg}{\emph{e.g.}}

\newcommand{\eq}{\leftarrow}

\graphicspath{{FIGS/}{PICTURES/}}

\def\ptitle{Heuristics for Packing Semifluids}
\def\pnumber{DCC-2016-01}
\def\pauthor{Jo\~ao Pedro Pedroso}

\trtitle{\ptitle}
\trnumber{\pnumber}
\trauthor{\pauthor}

\begin{document}
\mkcoverpage


\setcounter{page}{1}

\title{Heuristics for Packing Semifluids}



\author{Jo{\~a}o Pedro Pedroso}

\date{June 2015}

\maketitle

\begin{abstract}
  Physical properties of materials are seldom studied in the context of packing problems.  In this work we study the behavior of semifluids: materials with particular characteristics, that share properties both with solids and with fluids.  We describe the importance of some specific semifluids in an industrial context, and propose methods for tackling the problem of packing them, taking into account several practical requirements and physical constraints.  Although the focus of this paper is on the computation of practical solutions, it also uncovers interesting mathematical properties of this problem, which differentiate it from other packing problems.
  
  
\end{abstract}

\section{Introduction}
\label{sec:intro}

\begin{figure}
  \includegraphics[width=.5\textwidth]{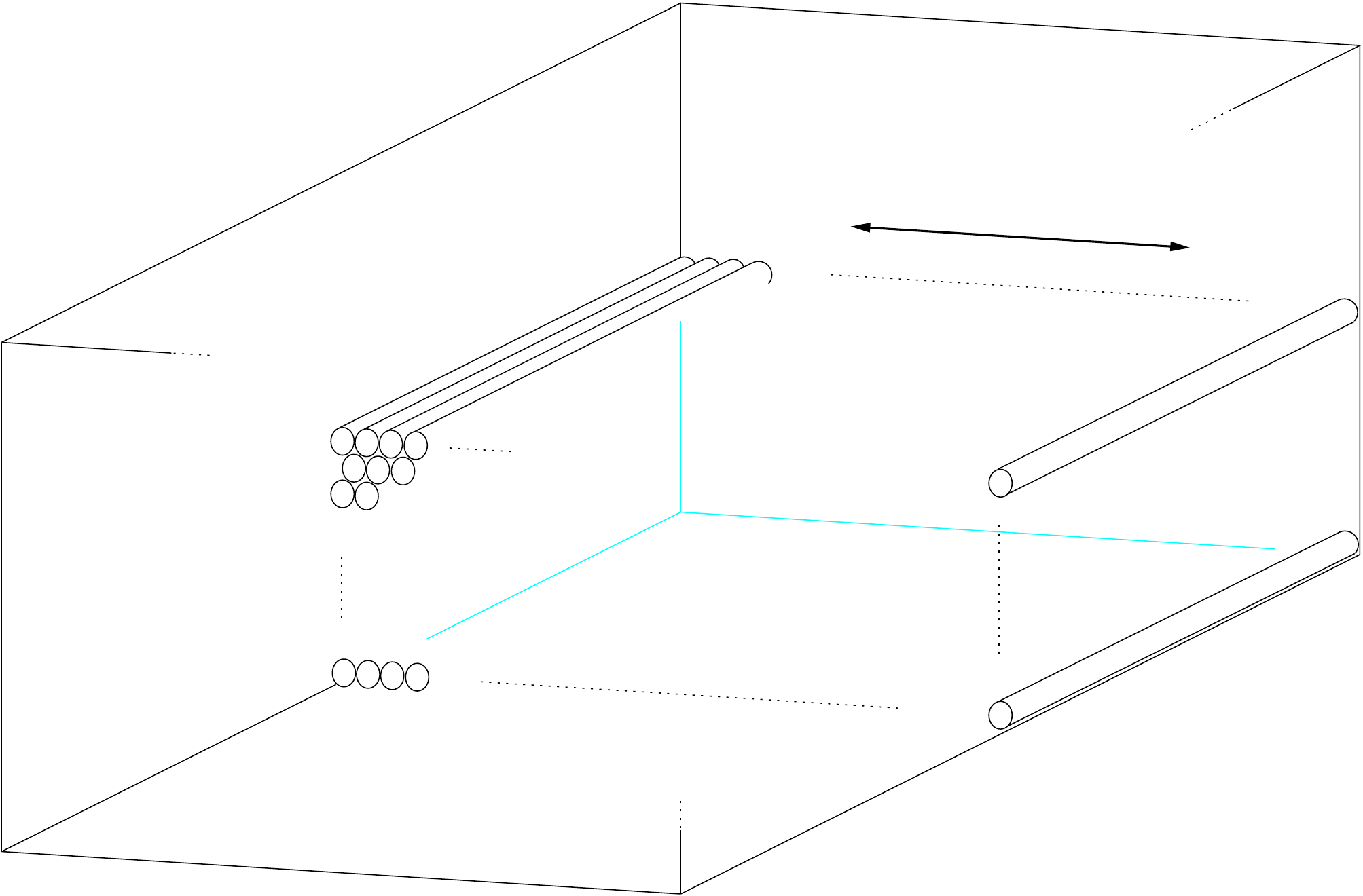}
  \hfill{}
  \scalebox{.35}{\begin{tikzpicture}[y=-1cm]

\draw[dotted,black] (20.95333,6.66667) -- (20.32,6.98444);
\draw[black] (12.7,5.08) -- (12.7,8.88889);
\draw[black] (2.54,17.78) -- (7.53778,15.28);
\draw[black] (22.86,5.71333) -- (22.86,13.33333) -- (12.7,18.41333);
\draw[black] (2.54,17.78) -- (12.7,18.41333);
\draw[black] (2.54,10.16) -- (2.54,17.78);
\draw[black] (12.7,18.41333) -- (12.7,17.46222);
\draw[dotted,black] (12.7,16.98444) -- (12.7,17.46222);
\draw[black] (2.54,10.16) -- (5.08,10.31778);
\draw[dotted,black] (5.08,10.31778) -- (5.72667,10.35556);
\draw[black] (2.54,10.16) -- (12.7,5.08) -- (22.86,5.71333) -- (20.95333,6.66667);
\draw[semithick,arrows=triangle 45-] (12.7,9.84222) -- (12.7,12.7);
\draw[semithick,arrows=-triangle 45] (12.7,12.7) -- (21.59111,13.25111);
\draw[semithick,arrows=-triangle 45] (12.7,12.7) -- (8.88889,14.60444);
\path (13.17556,10.47556) node[text=black,anchor=base west] {\fontsize{30.0}{36.0}\selectfont{}$z$};
\path (20.32,12.7) node[text=black,anchor=base west] {\fontsize{30.0}{36.0}\selectfont{}$y$};
\path (23.33556,9.68222) node[text=black,anchor=base west] {\fontsize{30.0}{36.0}\selectfont{}$H$};
\path (17.62,4.92) node[text=black,anchor=base west] {\fontsize{30.0}{36.0}\selectfont{}$W$};
\path (9.04667,13.81111) node[text=black,anchor=base west] {\fontsize{30.0}{36.0}\selectfont{}$x$};
\path (6.82444,7.30222) node[text=black,anchor=base west] {\fontsize{30.0}{36.0}\selectfont{}$D$};

\end{tikzpicture}%
} 
\caption{A container accommodating a semifluid: tubes (left); coordinate system used (right).}
\label{fig:pipes1}
\end{figure}
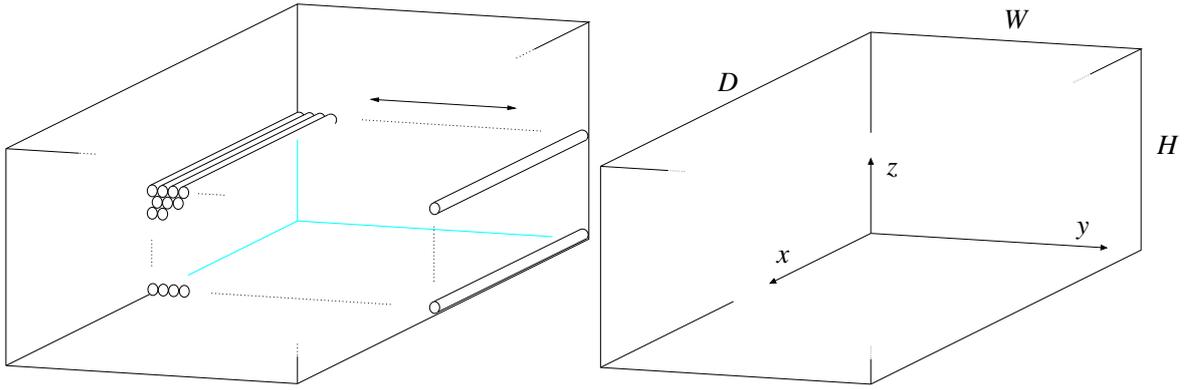

Semifluids are materials having characteristics of both fluids and solids.  In the context of this paper, we will consider materials that cannot flow in one direction, though they are fluid in the other directions.  As an example, consider tubes, which correspond to the industrial origin of this problem.  Placed in a container, they can flow in the directions perpendicular to their length, but \emph{not} in the direction of their length (see Figure~\ref{fig:pipes1}).  Assuming the tubes will be positioned perpendicularly to the Cartesian axes, depending on the direction of their placement they will flow either in the $x$ or in the $y$ dimension.  Pipes, having positive radii, are imperfect semifluids, as they will not fully occupy the space available in the $z$ dimension; however, they approximate a perfect fluid as the radii becomes smaller.  We will consider that the material is a perfect semifluid, and hence the volume occupied is constant and divisible.

This paper describes several possibilities for packing semifluids in a container, and presents heuristics for the variant which closer corresponds to an industrial application.

\section{Problem description}
\label{sec:problem}

Even though packing problems may be generalized into a single problem, they are usually divided in two categories: minimizing the number of bins, and maximizing the load to pack in a bin (see, \eg, \cite{Baldi20121205}).  Given an index set $\setS$ of semifluid items, with each item $i$ characterized by a fixed length $\ell_i$ and a volume $v_i$, and dimensions $D, W, H$ of containers, these two variants for the problem of packing a semifluid are:
\begin{enumerate}
\item bin packing variant: find the minimum number of containers to accommodate all the items;
\item knapsack variant: given, additionally, a value $w_i$ for the available volume $v_i$ of each item $i$, find the packing of maximum value that can be inserted in a container.
\end{enumerate}
In this paper we will focus on the knapsack variant.

\subsection{Semifluid packing problems}

\begin{figure}
  \scalebox{.25}{\begin{tikzpicture}[y=-1cm]

\draw[black] (11.21833,16.61583) ellipse (0.07408cm and 0.11642cm);
\draw[dotted,black] (20.955,6.6675) -- (20.32,6.985);
\draw[black] (12.7,5.08) -- (12.7,8.89);
\draw[cyan] (12.7,12.7) -- (2.54,17.78);
\draw[cyan] (12.7,12.7) -- (22.86,13.335);
\draw[cyan] (12.7,8.89) -- (12.7,12.7);
\draw[black] (2.54,17.78) -- (12.7,18.415);
\draw[black] (2.54,10.16) -- (2.54,17.78);
\draw[black] (12.7,18.415) -- (12.7,17.78);
\draw[dotted,black] (12.7,17.30375) -- (12.7,17.78);
\draw[black] (2.54,10.16) -- (5.08,10.31875);
\draw[dotted,black] (5.08,10.31875) -- (5.7277,10.35685);
\draw[black] (2.54,10.16) -- (12.7,5.08) -- (22.86,5.715) -- (20.955,6.6675);
\draw[semithick,black] (2.54,16.51) -- (2.54,17.78) -- (10.16,18.25625) -- (10.16,16.98625) -- cycle;
\draw[semithick,black] (2.54,16.51) -- (12.7,11.43) -- (20.32,11.90625) -- (10.16,16.98625);
\draw[semithick,black] (10.16,18.25625) -- (20.32,13.17625) -- (20.32,11.90625);
\draw[semithick,arrows=triangle 45-triangle 45,black] (2.54,18.0975) -- (10.16,18.57375);
\draw[black] (8.63388,16.3195) -- (11.17388,16.47825);
\draw[black] (8.66563,16.5608) -- (11.20563,16.71955);
\draw[dotted,black] (7.99042,16.27082) -- (8.63812,16.30892);
\draw[dotted,black] (8.001,16.51423) -- (8.6487,16.55233);
\draw[black] (22.86,5.715) -- (22.86,13.335) -- (12.7,18.415);
\path (6.35,19.26167) node[text=black,anchor=base] {\fontsize{31.68}{38.016}\selectfont{}$\ell$};
\path (1.74625,17.30375) node[text=black,anchor=base west] {\fontsize{31.68}{38.016}\selectfont{}$b$};
\path (6.6675,13.97) node[text=black,anchor=base west] {\fontsize{31.68}{38.016}\selectfont{}$a$};

\end{tikzpicture}%
}  \hfill{}
  \scalebox{.25}{\begin{tikzpicture}[y=-1cm]

\draw[black] (8.8265,11.29242) ellipse (0.08467cm and 0.11642cm);
\draw[dotted,black] (20.955,6.6675) -- (20.32,6.985);
\draw[black] (12.7,5.08) -- (12.7,8.89);
\draw[cyan] (12.7,12.7) -- (2.54,17.78);
\draw[cyan] (12.7,12.7) -- (22.86,13.335);
\draw[cyan] (12.7,8.89) -- (12.7,12.7);
\draw[black] (2.54,17.78) -- (12.7,18.415);
\draw[black] (2.54,10.16) -- (2.54,17.78);
\draw[black] (12.7,18.415) -- (12.7,17.78);
\draw[dotted,black] (12.7,17.30375) -- (12.7,17.78);
\draw[black] (2.54,10.16) -- (5.08,10.31875);
\draw[dotted,black] (5.08,10.31875) -- (5.7277,10.35685);
\draw[black] (2.54,10.16) -- (12.7,5.08) -- (22.86,5.715) -- (20.955,6.6675);
\draw[semithick,arrows=triangle 45-triangle 45,black] (18.415,15.875) -- (22.86,13.6525);
\draw[semithick,black] (18.415,11.7475) -- (18.415,15.5575) -- (8.255,14.9225) -- (8.255,11.1125);
\draw[black] (22.86,5.715) -- (22.86,13.335) -- (12.7,18.415);
\draw[semithick,black] (22.86,9.525) -- (22.86,13.335) -- (18.415,15.5575);
\draw[semithick,black] (12.7,8.89) -- (22.86,9.525) -- (18.415,11.7475) -- (8.255,11.1125) -- cycle;
\draw (10.4013,10.35685) -- (8.81592,11.15483);
\draw[dotted,black] (11.0236,10.0457) -- (10.3886,10.3632);
\draw (10.44152,10.6045) -- (8.85613,11.40248);
\draw[dotted,black] (11.0744,10.28488) -- (10.4394,10.60238);
\path (20.955,15.5575) node[text=black,anchor=base] {\fontsize{31.68}{38.016}\selectfont{}$\ell$};
\path (17.30375,8.89) node[text=black,anchor=base west] {\fontsize{31.68}{38.016}\selectfont{}$c$};
\path (23.01875,11.43) node[text=black,anchor=base west] {\fontsize{31.68}{38.016}\selectfont{}$d$};

\end{tikzpicture}%
}
\caption{Two possibilities for accommodating a semifluid in a container.}
\label{fig:pipes2}
\end{figure}
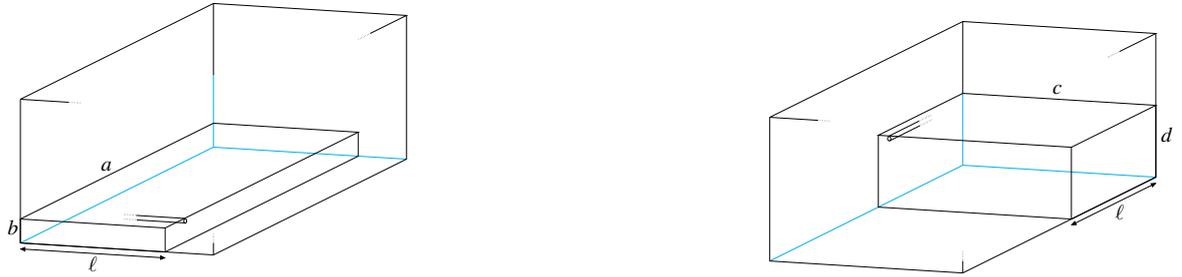

There are several possibilities for packing a semifluid orthogonally in a container, as shown in Figure~\ref{fig:pipes2}.  Both the length $\ell$ (corresponding to the length of the tubes) and the volume $v$ occupied by the semifluid are constant; in this figure, this means that $a \times b \times \ell = c \times d \times \ell = v$.  Assuming that, except the container itself, there are no walls, a semifluid will take on all the available horizontal space in the direction where it freely flows.  In the case presented, $a$ would take the depth $D$ of the container, and $c$ would take its width $W$, and hence the corresponding heights are $b = \frac{v}{D \ell}$ and  $d = \frac{v}{W \ell}$.  
After a semifluid is placed, others may be put on top of it, but they must not protrude (as detailed next).  Hence, one may think of the space above a semifluid as a ``container'', which can be filled up with the same rules as the original container; in this sense, this is a recursive problem.

\begin{figure}[t!]
  \centering
  \begin{center}
    \includegraphics[width=.45\textwidth]{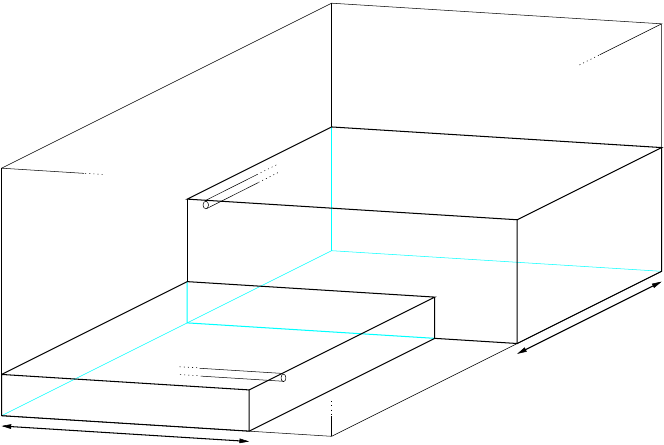}
    \hfill
    \includegraphics[width=.45\textwidth]{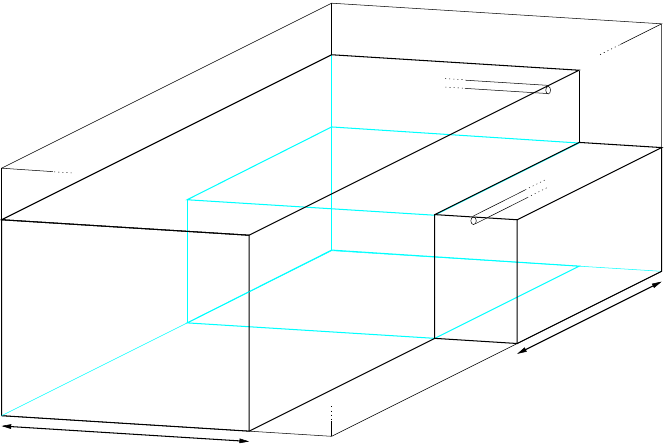}
  \end{center}
  \caption{Packing a semifluid without overflowing another previously packed (left), and overflowing it (right).}
  \label{fig:pipes3}
\end{figure}

Depending on the application, it may be allowed or not that, when packing a semifluid, it overflows others previously packed, as illustrated in Figure~\ref{fig:pipes3}.  In general, allowing overflow makes packing solutions more difficult to implement in practice, and brings the problem more difficult to tackle; overflow will not be considered here.
We will focus on packing semifluids by positioning the fixed dimension parallel to the $x$ axis, as shown in Figure~\ref{fig:pipes1}.  This is the relevant variant when the container must be loaded from a lateral door at $x=D$: if the semifluids were rotated and be placed along the $y$ axis, they would flow out of the door.  

An important, practical packing rule restricts what can be placed on top of what.  Indeed, for cargo stability and for facilitating loading, it is usually acceptable that shorter tubes are placed on top of longer tubes, but not the inverse; more precisely, there must be no holders protruding with respect to holders below them.

In semifluid packing, any fraction of an item's available volume may be packed; this is major difference with respect to other packing problems.

We call the problem of maximizing the value of semifluids packed in the container in these conditions the \emph{basic semifluid packing problem}.

\subsection{Background}
\label{sec:background}

Three-dimensional packing has recently been studied under several different perspectives; a recent survey can be found in~\cite{Crainic2012}.  The problem of allocating a given set of three-dimensional rectangular items to the minimum number of identical finite bins without overlapping has been addressed with tabu search in~\cite{Lodi2002410}: items are packed in several layers, the floor of the container being the first.  A heuristic method for the situation where there is no requirement for packed boxes to form flat layers, keeping track of empty space seen from different perspectives and using a look-ahead scheme for positioning, is presented in~\cite{Lim2003471}.  
However, the nature of the basic semifluid packing problem is rather different of these three-dimensional packing problems.  
As will be seen later, there is more similarity between our problem and two-dimensional cutting.  The most closely related problem is the orthogonal two-dimensional knapsack problem with guillotine patterns.  Methods for tackling this problem are often based on a discretization of possible positions for the rectangles in the Cartesian plane (see, \eg, \cite{Puchinger2007,Elsa2010,dolatabadi2012}).  A different approach is proposed in~\cite{fekete2007}, providing an exact algorithm for higher-dimensional orthogonal packing; the algorithm is based on bounding procedures which make use of dual feasible functions, within a tree search procedure.  With respect to these problems, semifluid packing has the property that it is not required to pack all the available volume of each item; in rectangle packing, this would correspond to being able to cut some of the rectangles at the time of packing.  Another difference between semifluid packing and previously studied problems concerns the requirement of no protuberance of items above others; this requirement is naturally respected in two-staged guillotine cuts, but usually is not enforced in general guillotine patterns.

To the best of our knowledge, basic semifluid packing or equivalent problems have not been studied before.  


\subsection{Mathematical model}

We are not aware of previous attempts to formulate the semifluid packing problem as a mathematical optimization model, but there are some related problems.  Integer programming models for two-dimensional two-stage bin packing problem have been proposed 
in~\cite{Lodi2004} and extended by~\cite{Puchinger2007} to the three-stage problem. In both cases, decision variables are related to the assignment of the items to bins, stripes or stacks.  Models for the related cutting stock problem, providing better linear relaxation bounds, are presented in~\cite{Elsa2010}, where a set of small rectangular items of given sizes is to be cut from a set of larger rectangular plates, in such a way that the total number of used plates is minimized.  Despite some similarities, none of these models is adequate for our problem, mainly for two reasons: in semifluid packing the number of stages is in general much larger, and items may be partially assigned to a position (\ie, a position may hold a fraction of the available volume of an item).

The formulation proposed next is not compact, as it requires an exponential number of variables; however, it hopefully conveys the characteristics of the problem.  For the sake of clarity, we start with a simplified model, and later describe how it could be extended to the general case; the simplification consists of assuming that only one stack of each item is allowed on each layer.  A layer, in this context, is either the floor of the container or the space above a previously packed item.  Figures \ref{fig:example} and~\ref{fig:instances} may be of help for visualizing the model.  

The first set of binary variables indicates which items are packed in the first layer: $y_i=1$ if item $i$ is packed directly on the container, $y_i=0$ otherwise.  To each variable $y_i$ there is a corresponding continuous variable $0 \leq x_i \leq 1$ which represents the fraction of item $i$ being packed at this place.  Before introducing more variables, let us specify a constraint related to the length $D$ of the container, which limits the length of items packed in this layer:
\begin{alignat*}{27}
  & \sum_i \ell_i y_i \leq D.               
\end{alignat*}
Variable $x_i$ may be positive only if $y_i=1$:
\begin{alignat*}{27}
  & x_i \leq y_i, && \quad \forall i.
\end{alignat*}
The height of the first layer is limited by the height of the container:
\begin{alignat*}{27}
  & h_i x_i \leq H, && \quad \forall i,      
\end{alignat*}
where $h_i = v_i/W$ is the total height that item $i$ would take on a container of width~$W$ (this will be later be replaced by a stronger constraint).

We now introduce variables concerning the placement of items $j$ on the second layer, \ie, directly above some previously packed item~$i$.  Variables $y_{ij}$ are the indicators for this, and the corresponding $x_{ij}$ represent the fraction of $j$ packed at this place.  The solution must, therefore, observe:
\begin{alignat*}{27}
  & y_{ij} \leq y_{i}, && \quad \forall i,j,\\
  & x_{ij} \leq y_{ij}, && \quad \forall i,j,\\
  & \sum_{j} \ell_{ij} y_{ij} \leq \ell_{i} y_{i}, && \quad \forall i,        
\end{alignat*}
where the last constraint limits the length of items placed directly above item~$i$.  For each pair $i,j$ the height of the corresponding stack is limited to the height of the container:
\begin{alignat*}{27}
  & h_i x_i + h_j x_{ij} \leq H, && \quad \forall i,j.                 
\end{alignat*}
The fraction of $i$ used in the two first layers is limited by one  (this and the previous constraints will later be extended):
\begin{alignat*}{27}
  & 0 \leq x_{i} + \sum_{j} x_{ji} \leq 1.                             
\end{alignat*}

We now have all the components to complete the model, by extending the number of layers.  Notice that, as layers cannot protrude and items with identical length should be placed by decreasing value, there may be at most $N$ layers, where $N$ is the number of items.  We may assume that the items are reversely ordered by length, \ie, $\ell_1 \geq \ell_2 \geq \ldots \geq \ell_N$; this allows us to define variables with indices $i,i'$ only for $i'>i$.  Notice also that the number of indices indicates the level at which the item corresponding to a variable is being packed:
variables for layer $1\leq K\leq N$ will have $K$ indices $i,j,\ldots,m,n$, with $i < j < \ldots < m < n$.
The entire model is presented in Figure~\ref{fig:model}.

\begin{figure}
  \centering
  \begin{normalsize}
\begin{alignat}{27}
  \maxim \quad & \sum_i w_i \tilde{x}_i  \label{eq:obj}
\end{alignat}
\raggedright{subject to:}
\begin{alignat}{27}
  & \tilde{x}_i = x_{i} + \sum_{j} x_{ji} + \sum_{k,j} x_{kji} + \ldots + \sum_{n,\ldots,j} x_{n\ldots ji}, \quad &&  \forall i \label{eq:totalx}\\[3mm]
  & \sum_i \ell_i y_i \leq D                 \label{eq:D}\\
  & \sum_{j} \ell_{ij} y_{ij} \leq \ell_{i} y_{i}, &&  \forall i   \nonumber\\
  & \ldots \nonumber\\
  & \sum_{n} \ell_{ij\ldots mn} y_{ij\ldots mn} \leq \ell_{ij\ldots m} y_{ij\ldots m}, &&  \forall i,j,\ldots,m  \label{eq:Dn}\\[3mm]
  & x_i \leq y_i, &&  \forall i          \label{eq:x1}\\
  & x_{ij} \leq y_{ij}, &&  \forall i,j \nonumber\\
  & \ldots \nonumber\\
  & x_{ij\ldots n} \leq y_{ij\ldots n}, &&  \forall i,j,\ldots,n \label{eq:xn}\\[3mm]
  & y_{ij} \leq y_{i}, &&  \forall i,j \label{eq:y1} \\
  & \ldots \nonumber\\
  & y_{ij\ldots mn} \leq y_{ij\ldots m}, &&  \forall i,j,\ldots,n  \label{eq:yn}\\[3mm]
  & h_i x_i + h_j x_{ij} + \ldots + h_n x_{ij\ldots n} \leq H, &&  \forall i,j,\ldots, n \label{eq:H}\\[3mm]
  & 0 \leq \tilde{x}_i \leq 1,  && \forall i \label{eq:beginvar}\\
  & 0 \leq x_{i} \leq 1, && \forall i \nonumber\\
  & \ldots \nonumber\\
  & 0 \leq x_{ij\ldots n} \leq 1, && \forall i,j,\ldots, n \nonumber\\[3mm]
  & y_{i} \in \{0,1\}, && \forall i  \nonumber\\
  & \ldots \nonumber\\
  & y_{ij\ldots n} \in \{0,1\}, && \forall i,j,\ldots, n  \label{eq:endvar}
\end{alignat}    
  \end{normalsize}
  \caption{Mathematical optimization model.}
  \label{fig:model}
\end{figure}

Equations~(\ref{eq:totalx}) determines the total quantity of item $i$ packed, which allows determining the total value packed in~(\ref{eq:obj}).  
Constraints (\ref{eq:D}) to~(\ref{eq:Dn}) guarantee that the total length of what is packed on top of the container, or of a packed item, does exceed the respective lengths.
Constraints (\ref{eq:x1}) to~(\ref{eq:xn}) allow a positive quantity of an item to packed only if the corresponding indicator variable is equal to~1.  
Constraints (\ref{eq:y1}) to~(\ref{eq:yn}) allow packing only on top of previously packed items.
Inequalities (\ref{eq:H}) determines the height of all stacks, and limits it to the height of the container.
Finally, (\ref{eq:beginvar}) to~(\ref{eq:endvar}) define the domain for each of the variables.

This model is rather clumsy, but it is not yet complete: it does not take into account the possibility of packing several stacks of each item on a given layer.  For make the model complete one would have to create, for each layer and each compatible item, a number of variables equal to the number of times that item would fit in the layer, if it was packed alone.  It is obvious that direct usage of this model is implausible, except for a rather small number of items; realistic usage would require a column generation approach.

\section{Heuristic and complete search}
\label{sec:heur}

For solving the basic semifluid packing problem, we firstly propose a heuristic method --- which will later be improved --- for dividing a container into smaller parallelepipeds, which we call \emph{holders}.  Each holder has a fixed depth, determined by the length of the semifluid it will accommodate.  Due to the possibility for the semifluid to flow downwards, along the $z$ dimension, and also along the $y$ dimension, a semifluid will fully use the width of the physical container.  The height of a filled holder is determined either by the volume of its semifluid or by the height of the physical container; in the latter case, the semifluid left over will possibly be packed in a different holder.

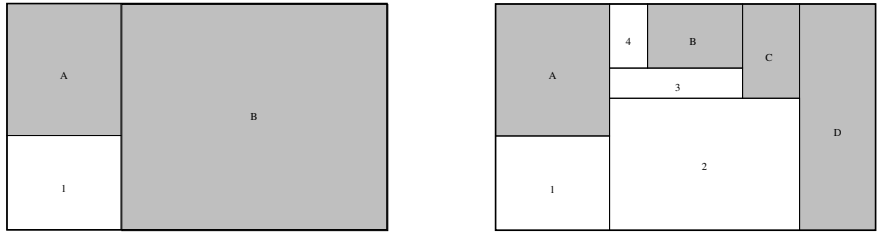
\begin{figure}
  \hfill{}
  \scalebox{.50}{\begin{tikzpicture}[y=-1cm]

\path[draw=black,semithick,fill=black!25] (0,0) rectangle (3,3.5);
\path[draw=black,semithick,fill=black!25] (3.02222,0.02222) rectangle (10.02222,6.02222);

\draw[semithick,black] (0,3.5) rectangle (3,6);
\draw[very thick,black] (0,0) rectangle (10,6);
\path (1.5,5) node[text=black,anchor=base] {\footnotesize{}1};

\path (1.5,2) node[text=black,anchor=base] {\footnotesize{}A};
\path (6.5,3.1) node[text=black,anchor=base] {\footnotesize{}B};

\end{tikzpicture}%
}  \hfill{}
  \scalebox{.50}{\begin{tikzpicture}[y=-1cm]

\path[draw=black,semithick,fill=black!25] (0,0) rectangle (3,3.5);
\path[draw=black,semithick,fill=black!25] (6.5,0) rectangle (8,2.5);
\path[draw=black,semithick,fill=black!25] (4,0) rectangle (6.5,1.7);
\path[draw=black,semithick,fill=black!25] (8,0) rectangle (10,6);

\draw[semithick,black] (0,3.5) rectangle (3,6);
\draw[semithick,black] (3,1.7) rectangle (6.5,2.5);
\draw[semithick,black] (3,0) rectangle (4,1.7);
\draw[semithick,black] (3,2.5) rectangle (8,6);
\draw[very thick,black] (0,0) rectangle (10,6);
\path (1.5,5) node[text=black,anchor=base] {\footnotesize{}1};
\path (4.8,2.3) node[text=black,anchor=base] {\footnotesize{}3};
\path (5.5,4.4) node[text=black,anchor=base] {\footnotesize{}2};
\path (3.5,1.1) node[text=black,anchor=base] {\footnotesize{}4};

\path (9,3.5) node[text=black,anchor=base] {\footnotesize{}D};
\path (7.2,1.5) node[text=black,anchor=base] {\footnotesize{}C};
\path (1.5,2) node[text=black,anchor=base] {\footnotesize{}A};
\path (5.2,1.1) node[text=black,anchor=base] {\footnotesize{}B};

\end{tikzpicture}%
}  \hfill{}
  \hfill{}
\caption{Section of a container through the $y=0$ plane: open holders (shaded) after placing one item (left), and after placing four items (right).}
\label{fig:example}
\end{figure}

In this situation, one may think of the packing process as a division of, say, the container's wall at $y=0$, into rectangles.  Each rectangle corresponds to the volume of a particular item when projected into the $y=0$ plane.  For example, consider the placement of a semifluid as in Figure~\ref{fig:pipes1}; a projection of the volume occupied is represented as a rectangle, alike~1 in the left diagram of Figure~\ref{fig:example}.  Upon placing this item, the container is divided into three partitions: one where item~1 is held (which is \emph{closed}, in the sense that it may not be used for other items), and the \emph{open} holders above (A) and besides the item~(B).  Upon placing three more items in this example, the open holders are A, B, C, D in the right diagram of Figure~\ref{fig:example}.

\subsection{Simple packing}
\label{packing}

A heuristic method for packing semifluids in these conditions can hence be though of as the process of choosing an item to pack, and an open holder for putting it (if some is available).  For a semifluid of length $\ell$, candidate holders $j$ must have depth $D_j \geq \ell$.  If the volume of a semifluid does not completely fit in the selected holder, the full height of the holder will be used (as for item 4 in the right diagram of Figure~\ref{fig:example}), and the remaining fluid is left to (possibly) pack later.

Given the characteristics of this problem, one might think of adapting known heuristics for bin packing and knapsack problems, as has been done for the two-dimensional knapsack problem (see, \eg,~\cite{coffman1980,dolatabadi2012}); however, the geometric constraint forbidding longer lengths on top of shorter leads to possibly unexpected performance, as we will see shortly.  Several alternative heuristic rules have been tried:
\begin{enumerate}
\item Best fit (BF): select the item/holder pair $(i,j)$ which leads to the minimum difference $D_j - \ell_i$, \ie, which leads to minimum currently unused space along~$x$;
\item Longest item first, first fit (LFF): select the longest item that can be packed in some open holder (\ie, item $i$ with largest $\ell_i$ for which there exists a holder $j$ such that $D_j - \ell_i \geq 0$), and insert it in the last open holder where it fits;
\item Longest item first, best fit (LBF): as LFF, but select the \emph{smallest} open holder in which the item fits;
\item Worthiest item first, first fit (WFF): as LFF, but select most valuable items (per unit volume) first;
\item Worthiest item first, best fit (WBF): as LBF, but select most valuable items first.
\end{enumerate}

\begin{algorithm}[htbp]
  \begin{footnotesize}
    \DontPrintSemicolon
    \SetKwFunction{algo}{algo}\SetKwFunction{pack}{pack}
    \SetKwFunction{algo}{algo}\SetKwFunction{h}{h}
    \KwData{instance:
      \begin{itemize}
      \item set $\setS$ of items to pack\;
      \item item's length $\ell_i$, volume $v_i$, and value $w_i$, $\forall i \in \setS$\;
      \item physical container's width $W$, height $H$,  and depth $D$;
      \end{itemize}
    }
    \KwResult{
      \begin{itemize}
      \item set of holders $\setH$ and their dimensions and position inside the container;
      \item for each item $i$, the set $x_i$ of holders where it is packed.
      \end{itemize}
    }
    \SetKwProg{myproc}{procedure}{}{}
    \SetKw{Break}{break}
    \myproc{\pack{$D, W, H, \setS, \ell, v, w$}}{
      $x_i \eq \{\}, \quad \forall i \in \setS$ \tcp*{initialize holders packing item $i$ as empty sets}
      $\setH \eq \{$holder with dimensions $D \times W \times H\}$ \tcp*{open main holder}
      \While{some item in $\setS$ fits in an holder in $\setH$}{
        $(i,j) \eq \h(\setS,\setH,\ell,v,w)$ \tcp*{heuristic choice of item $i$ and holder $j$}               \label{alg:rule}
        let $D_j, W_j, H_j$ be the current dimensions of holder $j$ \;
        $z \eq v_i / (\ell_i W_j)$ \;
        \If(\tcp*[f]{all volume of $i$ fits}){$z \leq H_j$}{
          $v_i \eq 0$\;
          $\setS \eq \setS \setminus \{i\}$ \;
          $(D_j,W_j,H_j) \eq (\ell_i,W_j,z)$ \tcp*{adjust $j$'s dimensions}
        }
        \Else{
          $v_i \eq (v_i - \ell_i W_j H_j)$ \tcp*{update volume of $i$ remaining unpacked}
          $(D_j,W_j,H_j) \eq (\ell_i,W_j,H_j)$ \tcp*{adjust $j$'s dimensions}
        }
        $x_i \eq x_i \cup \{j\}$ \tcp*{add $j$ to set of holders packing $i$}
        $\setH \eq \setH \setminus \{j\}$ \tcp*{remove $j$ from open holders}
        \If{$D_j > \ell_i$}{
          $\setH \eq \setH \cup \{$holder with dimensions $(D_j-\ell_i) \times W_j \times H_j\}$ \tcp*{open holder besides $j$}
        }
        \If{$H_j > z$}{
          $\setH \eq \setH \cup \{$holder with dimensions $\ell_i \times W_j \times (H_j-z)\}$ \tcp*{open holder on top of $j$}
        }
      }
      \Return{$x$}\;
    }
  \end{footnotesize}
  \caption{Simple heuristic method for packing semifluids.}
  \label{alg:pack}
\end{algorithm}

These rules are used in the heuristic method detailed in Algorithm~\ref{alg:pack}; we are abusing of notation, by allowing items and holders to be represented also by indices in their respective sets.  The algorithm returns a map associating each item to the set of holders that contain it (which is empty for items that are not packed).  The heuristic rule to be used is specified in line~\ref{alg:rule}, and holders are created accordingly in the subsequent lines.  The algorithm iterates as long as there is an open holder where some unpacked item fits.

The full description of the computational setup is deferred to Section~\ref{sec:results}; for the time being, we just present in Table~\ref{tab:pack} a comparison of the solutions obtained with these simple rules on a set of 3000 test instances.  We have counted the number of times that heuristic construction with a rule is \emph{strictly} better than with another, for all the combinations.
\begin{table}[h!tbp]
  \centering
  \caption{Comparison of simple rules for a data set of 3000 instances.  Left table: $n_{ij}$, the number of times rule $i$ was strictly better (\ie, found a better solution) than rule $j$.  Right table: $n_{ij} - n_{ji}$; positive values mean that rule on line $i$ is better for more instances than the rule in column $j$.}
  \label{tab:pack}
  \begin{tabular}{l|*{5}{r@{~~~~}}}
	& BF	& LFF	& LBF	& WFF	& WBF	\\\hline
BF	& 0	& 225	& 51	& 2526	& 2397	\\
LFF	& 338	& 0	& 101	& 2525	& 2396	\\
LBF	& 338	& 237	& 0	& 2529	& 2404	\\
WFF	& 339	& 342	& 336	& 0	& 91	\\
WBF	& 398	& 401	& 391	& 1737	& 0	\\
  \end{tabular}
~~~~~
  \begin{tabular}{l|*{5}{r@{~~~~}}}
	& BF	& LFF	& LBF	& WFF	& WBF	\\\hline
BF	& 0	& -113	& -287	& 2187	& 1999	\\
LFF	& 113	& 0	& -136	& 2183	& 1995	\\
LBF	& 287	& 136	& 0	& 2193	& 2013	\\
WFF	& -2187	& -2183	& -2193	& 0	& -1646	\\
WBF	& -1999	& -1995	& -2013	& 1646	& 0	\\
  \end{tabular}
\end{table}
The results obtained are rather surprising: rules based on the value of the items, very effective for the knapsack problem, are clearly outclassed by rules based on the length of the semifluid.  The simple rule of selecting the longest semifluid, independently of its value, and placing it in the open holder that leads to less used space along the $x$ axis (LBF) has generated the best results.  This is the heuristic rule selected for comparison with more elaborate methods.

\subsection{Local ascent}
\label{sec:local}

The previous packing algorithm can be easily extended to encompass local ascent, as proposed in Algorithm~\ref{alg:ascent}.  The idea is very simple: after finding a packing with the previous heuristics, attempt another construction forbidding items packed in the current solution, one at a time.  As soon as an improving solution is found, it is adopted as incumbent (\emph{first-improve}).  This process stops when all the neighbors of the current solution have been attempted, and they all lead to inferior solutions.

\begin{algorithm}[htbp]
  \begin{footnotesize}
    \DontPrintSemicolon
    \SetKwFunction{algo}{algo}\SetKwFunction{ascent}{ascent}
    \SetKwFunction{algo}{algo}\SetKwFunction{pack}{pack}
    \SetKwProg{myproc}{procedure}{}{}
    \SetKw{True}{true}
    \SetKw{False}{false}
    \SetKw{Break}{break}
    \myproc{\ascent{$D, W, H, \setS, \ell, v, w$}}{
      $x \eq \pack(D, W, H, \setS, \ell, v, w)$ \;
      let $\setI$ be the set of items packed in $x$ \;
      $\setT \eq \{\}$\;
      \Repeat{not improved}{
        improved = \False\;
        \For{$i \in \setI \setminus \setT$}{
          $\setT \eq \setT \cup \{ i\}$ \;
          $x' \eq \pack(D, W, H, \setS\setminus\{i\}, \ell, v, w)$ \;
          \If{value of $x'$ is greater than value of $x$}{
            $x \eq x'$ \;
            let $\setI$ be the set of items packed in $x$ \;
            improved = \True \;
            \Break \;
          }
        }
      }
      \Return{$x$}\;
    }
  \caption{Local ascent for packing semifluids.}
  \label{alg:ascent}
\end{footnotesize}
\end{algorithm}

This method is simple, and obviously finds a solution which is at least as good as that of Algorithm~\ref{alg:pack}.  As local ascent is still very fast, it is suitable for demanding situations (\eg, interactive processes).

\subsection{Complete search}
\label{sec:tree}

There are two reasons why the previous methods may be unsatisfactory.  The first reason concerns some rare, small instances for which a better solution can easily be found by inspection; the second reason concerns proving that the solution found is optimal.  We next propose some variants for doing complete search, based on tree search.

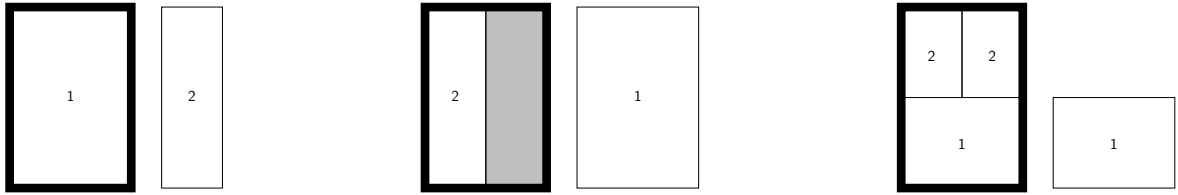
\begin{figure}
  \scalebox{.4}{\begin{tikzpicture}[y=-1cm]

\draw[semithick,black] (5,0) rectangle (7,6);
\draw[line width=8.1bp,black] (0,0) rectangle (4,6);
\path (6,3.1) node[text=black,anchor=base] {\Large\textsf{2}};
\path (2,3.1) node[text=black,anchor=base] {\Large\textsf{1}};

\end{tikzpicture}%
}  \hfill{}
  \scalebox{.4}{\begin{tikzpicture}[y=-1cm]

\path[draw=black,semithick,fill=white!75!black] (2,0) rectangle (4,6);

\draw[semithick,black] (5,0) rectangle (9,6);
\draw[semithick,black] (0,0) rectangle (2,6);
\draw[line width=8.1bp,black] (0,0) rectangle (4,6);
\path (1,3.1) node[text=black,anchor=base] {\Large\textsf{2}};
\path (7,3.1) node[text=black,anchor=base] {\Large\textsf{1}};

\end{tikzpicture}%
}  \hfill{}
  \scalebox{.4}{\begin{tikzpicture}[y=-1cm]

\draw[semithick,black] (0,0) rectangle (2,3);
\draw[semithick,black] (5,3) rectangle (9,6);
\draw[semithick,black] (2,0) rectangle (4,3);
\draw[line width=8.1bp,black] (0,0) rectangle (4,6);
\path (3,1.8) node[text=black,anchor=base] {\Large\textsf{2}};
\path (1,1.8) node[text=black,anchor=base] {\Large\textsf{2}};
\path (2,4.7) node[text=black,anchor=base] {\Large\textsf{1}};
\path (7,4.7) node[text=black,anchor=base] {\Large\textsf{1}};

\end{tikzpicture}%
}
\caption{An instance for which complete search does not find the optimum (shown in the rightmost diagram).  A vertical section of the container is represented with a bold line, and the item left over is shown beside it.}
\label{fig:subopt}
\end{figure}

Let us start with a caveat.  In the packing process we are considering, division of the semifluid occurs only when it does not fit vertically, and the amount left is possibly packed in another holder.  However, it may be optimal to fill only a part of the available amount of a semifluid.  This case is illustrated in Figure~\ref{fig:subopt}; if item 2 is more valuable than 1, it would be optimal to fill all the volume of item 2 over a part of 1, and leave the remaining 1 unpacked, as shown in the rightmost diagram.  However, visited solutions in a complete tree search are only the leftmost and the one in the center; hence, an \emph{``optimum''} for tree search many not be truly optimal for the original problem.

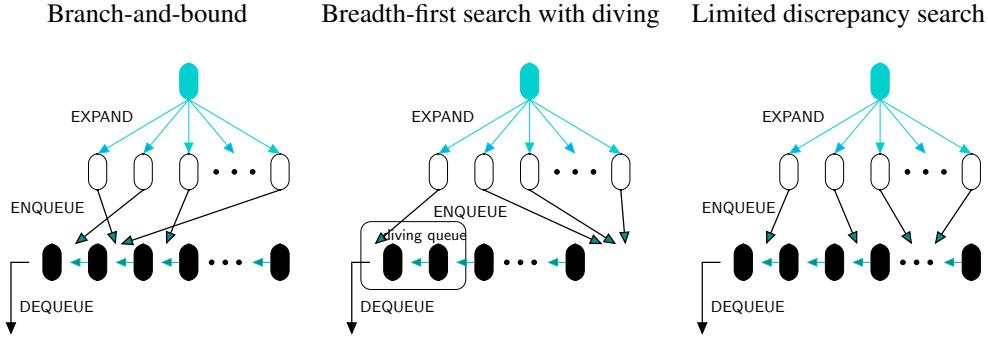
\begin{figure}[h!tbp]
  \centering
    \begin{tabular}{ccc}
    Branch-and-bound & Breadth-first search with diving & Limited discrepancy search\\
    ~\\
\scalebox{.60}{%
\begin{tikzpicture}[y=-1cm]

\definecolor{fillColor}{rgb}{0,0.56471,0.56471}
\draw[thick,fill=fillColor,arrows=-triangle 45] (8,3.8) -- (8.4,5);
\draw[thick,fill=fillColor,arrows=-triangle 45] (9,3.8) -- (7.5,5);
\draw[thick,fill=fillColor,arrows=-triangle 45] (10,3.8) -- (9.5,5);
\draw[thick,fill=fillColor,arrows=-triangle 45] (12,3.8) -- (8.5,5);

\fill[draw=black] (10.6,3.4) circle (0.05111cm);
\fill[draw=black] (11,3.4) circle (0.05111cm);
\fill[draw=black] (11.4,3.4) circle (0.05111cm);
\fill[draw=black] (10.5,5.4) circle (0.05111cm);
\fill[draw=black] (10.8,5.4) circle (0.05111cm);
\fill[draw=black] (11.1,5.4) circle (0.05111cm);
\filldraw[rounded corners=6.3bp] (7.2,5.8) rectangle (6.8,5);
\draw[rounded corners=6.3bp] (8.2,3.8) rectangle (7.8,3);
\draw[rounded corners=6.3bp] (9.2,3.8) rectangle (8.8,3);
\draw[rounded corners=6.3bp] (10.2,3.8) rectangle (9.8,3);
\draw[rounded corners=6.3bp] (12.2,3.8) rectangle (11.8,3);
\definecolor{penColor}{rgb}{0,0.81569,0.81569}
\filldraw[rounded corners=6.3bp,penColor] (10.2,1.8) rectangle (9.8,1);
\path[draw=penColor,semithick,fill=cyan,arrows=-triangle 45] (10,1.8) -- (9,3);
\path[draw=penColor,semithick,fill=cyan,arrows=-triangle 45] (10,1.8) -- (11,3);
\filldraw[semithick,penColor,arrows=-triangle 45] (10,1.8) -- (12,3);
\filldraw[rounded corners=6.3bp] (8.2,5.8) rectangle (7.8,5);
\filldraw[rounded corners=6.3bp] (9.2,5.8) rectangle (8.8,5);
\filldraw[rounded corners=6.3bp] (10.2,5.8) rectangle (9.8,5);
\filldraw[rounded corners=6.3bp] (12.2,5.8) rectangle (11.8,5);
\draw[thick,arrows=-triangle 45,black] (6.5,5.4) -- (6.1,5.4) -- (6.1,7);
\path[draw=penColor,semithick,fill=cyan,arrows=-triangle 45] (10,1.8) -- (8,3);
\filldraw[semithick,penColor,arrows=-triangle 45] (10,1.8) -- (10,3);
\definecolor{penColor}{rgb}{0,0.6902,0.6902}
\path[draw=penColor,fill=fillColor,arrows=-triangle 45] (7.8,5.4) -- (7.4,5.4);
\path[draw=penColor,fill=fillColor,arrows=-triangle 45] (8.8,5.4) -- (8.4,5.4);
\path[draw=penColor,fill=fillColor,arrows=-triangle 45] (9.8,5.4) -- (9.4,5.4);
\path[draw=penColor,fill=fillColor,arrows=-triangle 45] (11.8,5.4) -- (11.4,5.4);
\path (8.1,2.3) node[text=black,anchor=base] {\textsf{EXPAND}};
\path (7.1,6.5) node[text=black,anchor=base] {\textsf{DEQUEUE}};
\path (6.9,4.3) node[text=black,anchor=base] {\textsf{ENQUEUE}};

\end{tikzpicture}%
}
    & 
\scalebox{.60}{%
\begin{tikzpicture}[y=-1cm]

\definecolor{fillColor}{rgb}{0,0.56471,0.56471}
\draw[thick,fill=fillColor,arrows=-triangle 45] (8,3.8) -- (6.6,5);
\draw[thick,fill=fillColor,arrows=-triangle 45] (9,3.8) -- (11.6,5);
\draw[thick,fill=fillColor,arrows=-triangle 45] (10,3.8) -- (11.9,5);
\draw[thick,fill=fillColor,arrows=-triangle 45] (12,3.8) -- (12.1,5);

\fill[draw=black] (10.6,3.4) circle (0.05111cm);
\fill[draw=black] (11,3.4) circle (0.05111cm);
\fill[draw=black] (11.4,3.4) circle (0.05111cm);
\fill[draw=black] (9.5,5.4) circle (0.05111cm);
\fill[draw=black] (9.8,5.4) circle (0.05111cm);
\fill[draw=black] (10.1,5.4) circle (0.05111cm);
\filldraw[rounded corners=6.3bp] (7.2,5.8) rectangle (6.8,5);
\draw[rounded corners=6.3bp] (8.2,3.8) rectangle (7.8,3);
\draw[rounded corners=6.3bp] (9.2,3.8) rectangle (8.8,3);
\draw[rounded corners=6.3bp] (10.2,3.8) rectangle (9.8,3);
\draw[rounded corners=6.3bp] (12.2,3.8) rectangle (11.8,3);
\definecolor{penColor}{rgb}{0,0.81569,0.81569}
\filldraw[rounded corners=6.3bp,penColor] (10.2,1.8) rectangle (9.8,1);
\path[draw=penColor,semithick,fill=cyan,arrows=-triangle 45] (10,1.8) -- (9,3);
\path[draw=penColor,semithick,fill=cyan,arrows=-triangle 45] (10,1.8) -- (11,3);
\filldraw[semithick,penColor,arrows=-triangle 45] (10,1.8) -- (12,3);
\filldraw[rounded corners=6.3bp] (8.2,5.8) rectangle (7.8,5);
\filldraw[rounded corners=6.3bp] (9.2,5.8) rectangle (8.8,5);
\draw[thick,arrows=-triangle 45,black] (6.5,5.4) -- (6.1,5.4) -- (6.1,7);
\filldraw[rounded corners=6.3bp] (11.2,5.8) rectangle (10.8,5);
\draw[rounded corners=6.3bp,black] (8.6,6) rectangle (6.3,4.5);
\path[draw=penColor,semithick,fill=cyan,arrows=-triangle 45] (10,1.8) -- (8,3);
\filldraw[semithick,penColor,arrows=-triangle 45] (10,1.8) -- (10,3);
\definecolor{penColor}{rgb}{0,0.6902,0.6902}
\path[draw=penColor,fill=fillColor,arrows=-triangle 45] (7.8,5.4) -- (7.4,5.4);
\path[draw=penColor,fill=fillColor,arrows=-triangle 45] (8.8,5.4) -- (8.4,5.4);
\path[draw=penColor,fill=fillColor,arrows=-triangle 45] (10.8,5.4) -- (10.4,5.4);
\path (8.1,2.3) node[text=black,anchor=base] {\textsf{EXPAND}};
\path (7.1,6.5) node[text=black,anchor=base] {\textsf{DEQUEUE}};
\path (7.7,4.9) node[text=black,anchor=base] {\textsf{diving queue}};
\path (8.7,4.4) node[text=black,anchor=base] {\textsf{ENQUEUE}};

\end{tikzpicture}%
}
    & 
\scalebox{.60}{%
\begin{tikzpicture}[y=-1cm]

\definecolor{fillColor}{rgb}{0,0.56471,0.56471}
\draw[thick,fill=fillColor,arrows=-triangle 45] (8,3.8) -- (7.5,5);
\draw[thick,fill=fillColor,arrows=-triangle 45] (9,3.8) -- (9.5,5);
\draw[thick,fill=fillColor,arrows=-triangle 45] (10,3.8) -- (10.7,5);
\draw[thick,fill=fillColor,arrows=-triangle 45] (12,3.8) -- (11.2,5);

\fill[draw=black] (10.6,3.4) circle (0.05111cm);
\fill[draw=black] (11,3.4) circle (0.05111cm);
\fill[draw=black] (11.4,3.4) circle (0.05111cm);
\fill[draw=black] (10.5,5.4) circle (0.05111cm);
\fill[draw=black] (10.8,5.4) circle (0.05111cm);
\fill[draw=black] (11.1,5.4) circle (0.05111cm);
\filldraw[rounded corners=6.3bp] (7.2,5.8) rectangle (6.8,5);
\draw[rounded corners=6.3bp] (8.2,3.8) rectangle (7.8,3);
\draw[rounded corners=6.3bp] (9.2,3.8) rectangle (8.8,3);
\draw[rounded corners=6.3bp] (10.2,3.8) rectangle (9.8,3);
\draw[rounded corners=6.3bp] (12.2,3.8) rectangle (11.8,3);
\definecolor{penColor}{rgb}{0,0.81569,0.81569}
\filldraw[rounded corners=6.3bp,penColor] (10.2,1.8) rectangle (9.8,1);
\path[draw=penColor,semithick,fill=cyan,arrows=-triangle 45] (10,1.8) -- (9,3);
\path[draw=penColor,semithick,fill=cyan,arrows=-triangle 45] (10,1.8) -- (11,3);
\filldraw[semithick,penColor,arrows=-triangle 45] (10,1.8) -- (12,3);
\filldraw[rounded corners=6.3bp] (8.2,5.8) rectangle (7.8,5);
\filldraw[rounded corners=6.3bp] (9.2,5.8) rectangle (8.8,5);
\filldraw[rounded corners=6.3bp] (10.2,5.8) rectangle (9.8,5);
\filldraw[rounded corners=6.3bp] (12.2,5.8) rectangle (11.8,5);
\draw[thick,arrows=-triangle 45,black] (6.5,5.4) -- (6.1,5.4) -- (6.1,7);
\path[draw=penColor,semithick,fill=cyan,arrows=-triangle 45] (10,1.8) -- (8,3);
\filldraw[semithick,penColor,arrows=-triangle 45] (10,1.8) -- (10,3);
\definecolor{penColor}{rgb}{0,0.6902,0.6902}
\path[draw=penColor,fill=fillColor,arrows=-triangle 45] (7.8,5.4) -- (7.4,5.4);
\path[draw=penColor,fill=fillColor,arrows=-triangle 45] (8.8,5.4) -- (8.4,5.4);
\path[draw=penColor,fill=fillColor,arrows=-triangle 45] (9.8,5.4) -- (9.4,5.4);
\path[draw=penColor,fill=fillColor,arrows=-triangle 45] (11.8,5.4) -- (11.4,5.4);
\path (8.1,2.3) node[text=black,anchor=base] {\textsf{EXPAND}};
\path (7.1,6.5) node[text=black,anchor=base] {\textsf{DEQUEUE}};
\path (6.9,4.3) node[text=black,anchor=base] {\textsf{ENQUEUE}};

\end{tikzpicture}%
}
  \end{tabular}
  
  \caption{Queueing methods: branch-and-bound (left), where nodes in the queue are sorted by their upper bound; breadth-first search with diving (center), where no information about about the nodes entering the queue is used (at each expansion, one node generated is the diving node); and limited discrepancy search (right), where nodes are sorted by discrepancy (at each expansion, nodes are generated in this order).}
  \label{fig:queue}
\end{figure}

Complete search is an extension of Algorithm~\ref{alg:pack} where, instead of considering only packing the item chosen by the heuristic rule in line~\ref{alg:rule}, we consider all the possibilities of placing available items in open containers; each of these possibilities leads to a new node in the search tree.  Notice that the branching factor is very large, and hence straightforward complete search is prohibitive even for small instances.  Next, we present three relevant tree search alternatives for dealing with this difficulty; a visual insight of the differences between them is provided in Figure~\ref{fig:queue}.

\subsubsection{Branch-and-bound}
\label{sec:bb}

Branch-and-bound (BB) is the standard method for searching a tree in optimization (see, \eg,~\cite{lawler66} for an early survey).  For a maximization problem, the comparison of an upper bound of the objective that can be reached from a given node, to a known lower bound of the objective, is used to eliminate from consideration parts of the search tree.   The best solution visited so far is commonly used as the lower bound.  In the case of the basic semifluid packing problem, an upper bound can be obtained by sorting the items by decreasing unit value, and filling the space still available in the container by this order, assuming no shape constraints (this is similar to the linear relaxation bound for the knapsack problem; see~\cite{martello1990}).  For a given partial solution, holders that cannot be filled due to having no unpacked items that fit inside them are withdrawn from the list of open holders;  their volume is subtracted from the space available when computing the corresponding upper bound.

Another important factor for having a reasonably effective branch-and-bound concerns avoiding symmetric, or otherwise equivalent solutions.  This is done with the following rules:
\begin{itemize}
\item items placed at the same horizontal level must have increasing indices in the set $\setS$ of semifluids to pack;
\item items placed on top of given item $i$ having the same length as $i$ cannot have a larger unit value than~$i$.
\end{itemize}

The main steps of the branch-and-bound algorithm are outlined in Algorithm~\ref{alg:bb} (see also Appendix~\ref{sec:data}).  The algorithm is based on the iteration over elements in a queue ($Q$) until it becomes empty.  Nodes whose upper bound is inferior to the objective value of the best known solution are discarded (line~\ref{alg:pruning}).  Branching is carried out in lines \ref{alg:branchS}--\ref{alg:branchE}.  As all the possible assignments of yet unpacked items to open holders must be considered, the main limitation of the algorithm concerns the large number of nodes added in these lines.

The algorithm has two parameters, limiting CPU time and the size of the queue.  The latter is used when restricting the number of open nodes is required for keeping memory usage acceptable; in such cases, we provide the possibility of removing a part of the queue (\emph{chopping}, lines \ref{alg:chopS}--\ref{alg:chopE}).  When this occurs, as well as when the time limit is reached, the solution returned may be not optimal.  In the experiment reported in the Section~\ref{sec:results}, the maximum number of nodes is set to infinity, making CPU time the only factor limiting the search.

\newcommand{\PP}{P}	
\newcommand{\SP}{S}	
\newcommand{\UB}{\textit{UB}}
\newcommand{\LB}{\textit{LB}}
\newcommand{\OPT}{\textit{OPT}}
\begin{algorithm}[htbp!]
  \SetKw{True}{true}
  \SetKw{False}{false}
  \SetKw{Continue}{continue}
  \DontPrintSemicolon

  create a queue $Q$ with one node (the root relaxation) \tcp*{Initialization}
  set upper bound $\UB \eq \infty$, lower bound $\LB \eq -\infty$, optimality flag $\OPT \eq \True$ \;

  \Repeat{$Q = \{\}$ or time limit has been reached}{
    select and remove from $Q$ node $k$ with largest $\UB$ \label{alg:pruning} \tcp*{Subproblem selection}
    \If(\tcp*[f]{Pruning and fathoming}){$\UB^k \leq \LB$ or no items fit in open holders}{
      \Continue
    }

    \ForEach(\tcp*[f]{Partitioning}){feasible assignment of unpacked items to open holders \label{alg:branchS}}{
      add new node $n$ to $Q$\;
      \If{$\LB^n > \LB$}{
        update $\LB \eq \LB^n$\; \label{alg:branchE}
      }
    }

    \While(\tcp*[f]{Chopping}){size of $Q$ is larger than the allowed limit \label{alg:chopS}}{
      remove from $Q$ node with smallest $\UB$\;
      $\OPT \eq \False$ \; \label{alg:chopE}    
    }
     \If(\tcp*[f]{Termination}){time limit has been reached} {
       $\OPT \eq \False$ \;
     }
  }
  \Return{solution that yielded $\LB$, with optimality flag $\OPT$}
  \caption{Main steps of the branch-and-bound algorithm.}\label{alg:bb}
\end{algorithm}


\subsubsection{Breadth-first search with diving}
\label{sec:bfs}

As the branching factor is very large, standard branch-and-bound may not be allowed the time and space to produce a good solution, even for relatively small instances.  Indeed, as will be seen in the next session, in a limited time the solution of branch-and-bound is often worse than that of the simple heuristics.  For overcoming this issue, several alternatives have been proposed in the literature; these are usually based on \emph{diving} (see, \eg, \cite{achterberg2008,pochet2006}).
We firstly propose what we call \emph{breadth-first with diving (BFD)}, which consists of the following:
\begin{enumerate}
\item Keep two search queues: the main queue $Q$ and the \emph{diving queue} $R$;
\item If $R$ is not empty, at the current iteration explore the last element added to this queue (\ie, explore $R$ in a last in, first out manner);
\item If $R$ is empty, at the current iteration explore the first element added to $Q$ (\ie, explore $Q$ in a first in, first out manner);
\item When creating children of the current node, append the one that corresponds to the heuristic rule (LBF) to the queue $R$, and the remaining children (generated by decreasing item length) to $Q$.
\end{enumerate}
Hence, $R$ is searched in a last in, first out fashion, corresponding to the order of the LBF heuristic rule (longest item first, best fit container); therefore, the first leaf visited is the LBF solution.  The exploration of $Q$ in a breadth-first (first in, first out) fashion introduces diversity in the search, which balances well with the intensive search of the dive; this is important for time-limited executions, where parts of the tree are left unexplored.  Furthermore, quickly finding solutions of good quality allows pruning more nodes in the search tree.  Notice that as long as the item list is initially sorted by length, we can generate new nodes to add to $Q$ without further sorting (however, sorting available items by value is required for computing the upper bound of a new node).

Diving does not interact well with the symmetry breaking rules: if the diving item was forbidden for avoiding symmetry, the first dive would be interrupted, and the corresponding heuristic solution would not be reached.  In order to assure that we reach that solution, rules for avoiding symmetry are not enforced during diving.

In our implementation of BFS we are using the bounds described in Section~\ref{sec:bb}, which in most cases allow pruning significant parts of the search tree.

\subsubsection{Limited discrepancy search}
\label{sec:lds}
Another alternative to standard branch-and-bound is \emph{limited discrepancy search (LDS)}, where the tree is searched by increasing order of the \emph{number of violations} of the heuristic rule, as proposed in \cite{harvey1995}.  This method has been attempted with the LBF heuristic rule, but the computation of discrepancy in this case requires sorting the moves available, using considerable computational time.  A better alternative is to base the search in the longest item first, first fit rule (LFF); this allows a very quick expansion of nodes at each iteration, and exploring much larger parts of the tree in a limited time.

As in the standard version of LDS, this method uses a parameter specifying the discrepancy level above which search is abandoned.  This usually allows adjusting the part of the tree that is explored to the resources available, as an alternative to simply interrupting the execution after a certain time has elapsed.  We acknowledge that better solutions are often found with such an adjustment, and that memory usage will make the search impractical for long-running executions without limiting discrepancy; however, for an easier comparison with the other methods, we have set the discrepancy limit to infinity.  Due to this choice, whenever LDS ends before reaching the limit CPU time, its solution is optimal.

In our implementation of LDS we are using the bounds described in Section~\ref{sec:bb}, which in most cases allow pruning significant parts of the search tree.

\section{Computational results}
\label{sec:results}

In order to assess the performance of the methods proposed, we have created a set of instances based on the characteristics of the real-world application.  Practical instances we are aware of are small, as is the number of different semifluid lengths (tubes are usually cut in standard lengths).  Instances with more than 20 items go beyond the application's requirement, but are useful for testing the behavior of the different algorithms.  Instances are classified into two main families:
\begin{itemize}
\item Easy instances: generated in such a way that in the optimum there are no items left unpacked; for these instances, an optimal solution completely occupying the container is known.
\item Hard instances: no optimum is known in advance; the volume of available items corresponds either to 100\% of the container (as for easy instances, though now it is unlikely that all items can be packed), or to 150\% of it.
\end{itemize}
The number of semifluids considered are 5, 10, 20, 50, and~100.  Some instances have just a few distinct item lengths, other have more diverse lengths.
For each combination of these characteristics, 100 different instances have been generated, totaling 3000 instances.  A visualization of instances from the easy and hard subsets, with corresponding optimal and heuristic solutions, is provided in Figure~\ref{fig:instances} (details on the instance generator are available in Appendix~\ref{sec:data}).

\begin{figure}[htbp]
  \centering
  \begin{center}
    \includegraphics[width=.49\textwidth,trim=110 40 100 40,clip=True]{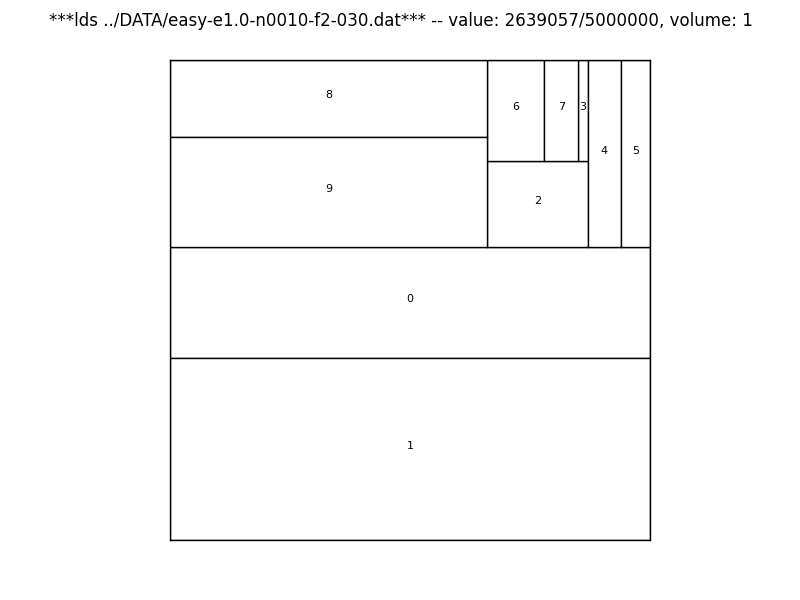}
    \hfill
    \includegraphics[width=.49\textwidth,trim=110 40 100 40,clip=True]{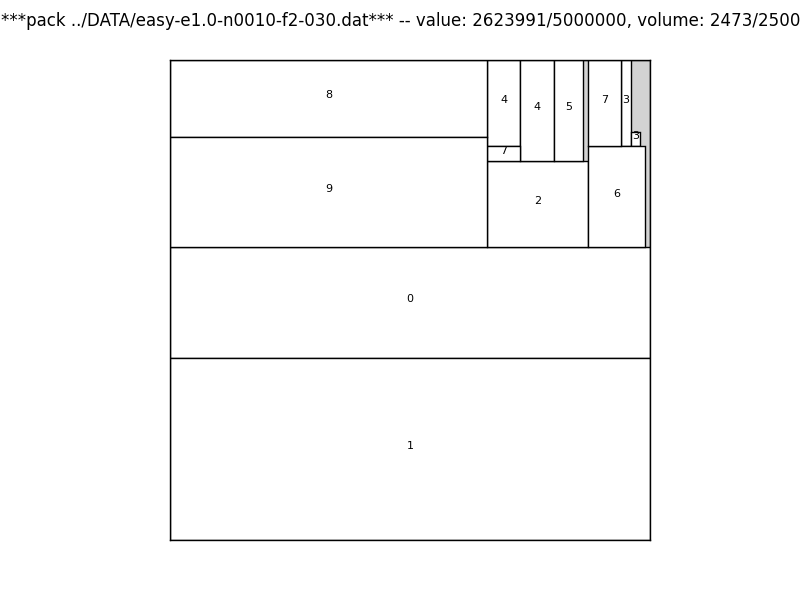}
  \end{center}
  \begin{center}
    \includegraphics[width=.49\textwidth,trim=110 40 100 40,clip=True]{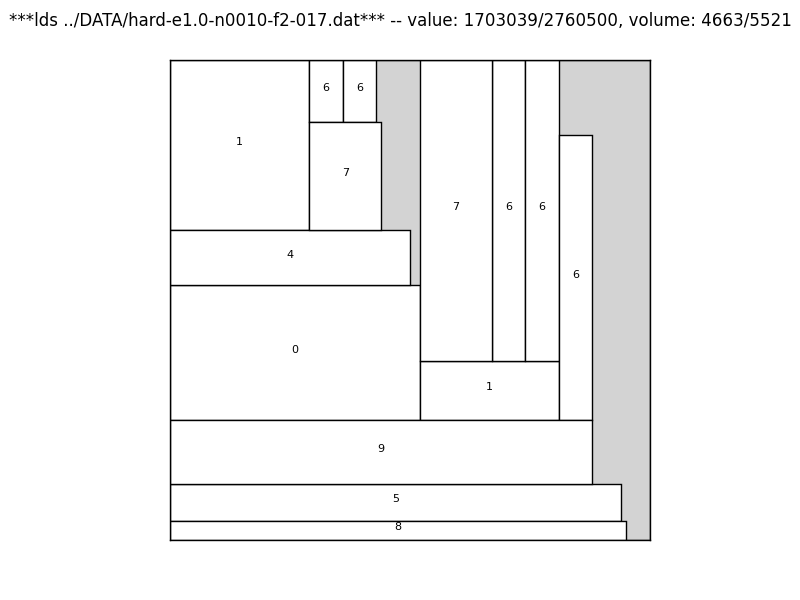}
    \hfill
    \includegraphics[width=.49\textwidth,trim=110 40 100 40,clip=True]{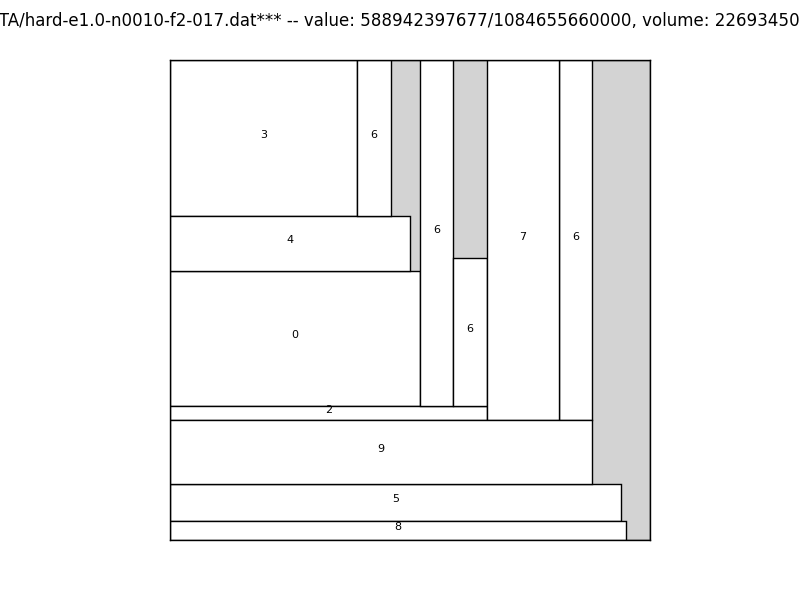}
  \end{center}
  \caption{An optimal solution (left), and a heuristic solution (right) for instances with ten items: an easy instance (top) and a hard instance (bottom).}
  \label{fig:instances}
\end{figure}

Our programs use exact arithmetic for all operations (hence, values in the instance files are written as fractions).  All the executions were limited to 60 seconds of CPU time, and both the maximum number of nodes and the discrepancy limit were set to infinity.

We start recalling the comparison among simple heuristics (Table~\ref{tab:pack}).  Having selected LBF, we now compare it to more elaborate methods in Table~\ref{tab:all}.  As expected, local ascent is always at least as good as LBF, being strictly superior for a massive share of instances.  As the CPU time limitation is rather severe, local ascent is also often better than tree search methods.  The best results overall have been obtained by limited discrepancy search.

\begin{table}[h!tbp]
  \centering
  \caption{Comparison of simple rule (LBF), local ascent (LA), and tree search --- standard branch-and-bound version (BB), breadth-first search with diving (BFD), and limited discrepancy search (LDS) --- for a data set of 3000 instances.  Left table: $n_{ij}$, the number of times method $i$ was strictly better (\ie, found a better solution) than method $j$.  Right table: $n_{ij} - n_{ji}$; positive values mean the method on line $i$ is better for more instances than the method in column $j$.}
  \label{tab:all}
  \begin{tabular}{l|*{7}{r@{~~~}}}
	& LBF	& LA	& BB	& BFS	& LDS	\\\hline
LBF	& 0	& 0	& 1627	& 306	& 42	\\
LA	& 2041	& 0	& 1744	& 849	& 187	\\
BB	& 909	& 633	& 0	& 108	& 101	\\
BFS	& 1520	& 1092	& 1895	& 0	& 329	\\
LDS	& 2007	& 1350	& 1915	& 1192	& 0	\\
  \end{tabular}
~~~~~
  \begin{tabular}{l|*{7}{r@{~~~}}}
	& LBF	& LA	& BB	& BFS	& LDS	\\\hline
LBF	& 0	& -2041	& 718	& -1214	& -1965	\\
LA	& 2041	& 0	& 1111	& -243	& -1163	\\
BB	& -718	& -1111	& 0	& -1787	& -1814	\\
BFS	& 1214	& 243	& 1787	& 0	& -863	\\
LDS	& 1965	& 1163	& 1814	& 863	& 0	\\
  \end{tabular}
\end{table}

\begin{figure}
  \centering
  \begin{center}
    \includegraphics[width=.75\textwidth,trim=30 10 70 0,clip=True]{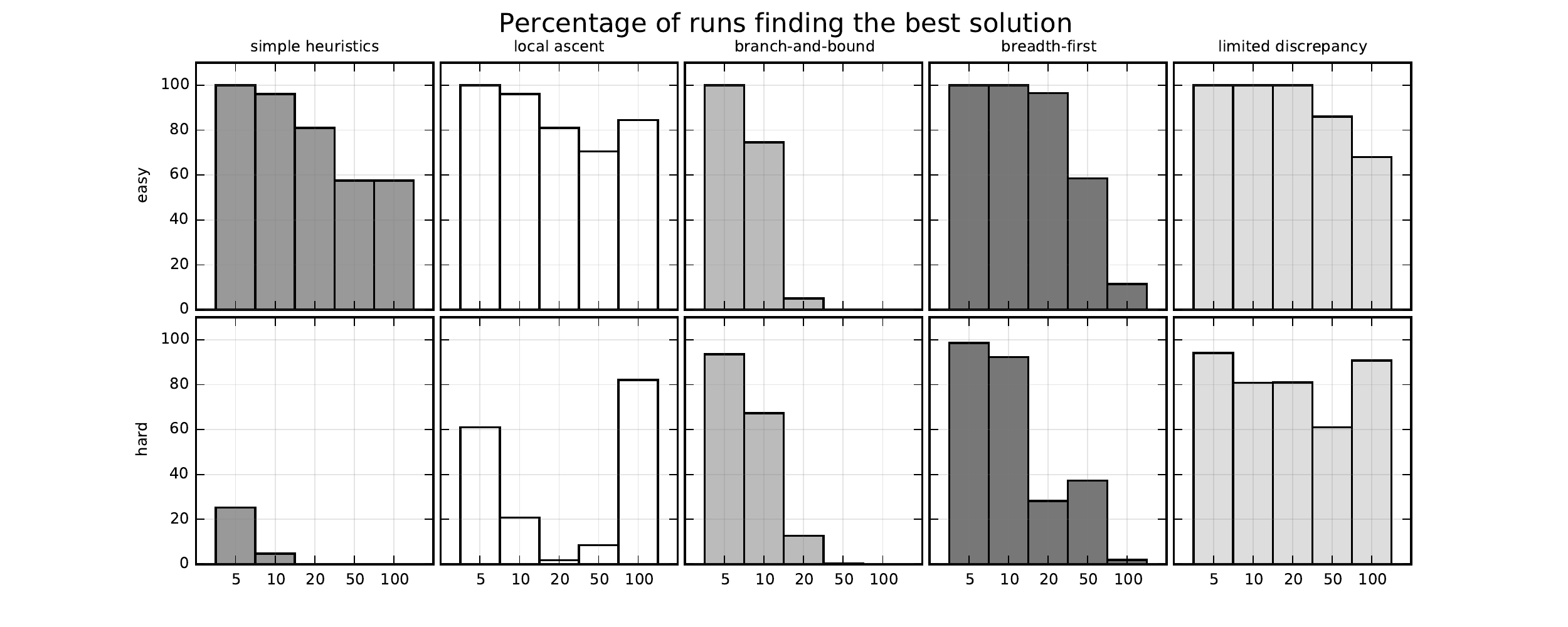}
    \includegraphics[width=.75\textwidth,trim=30 10 70 0,clip=True]{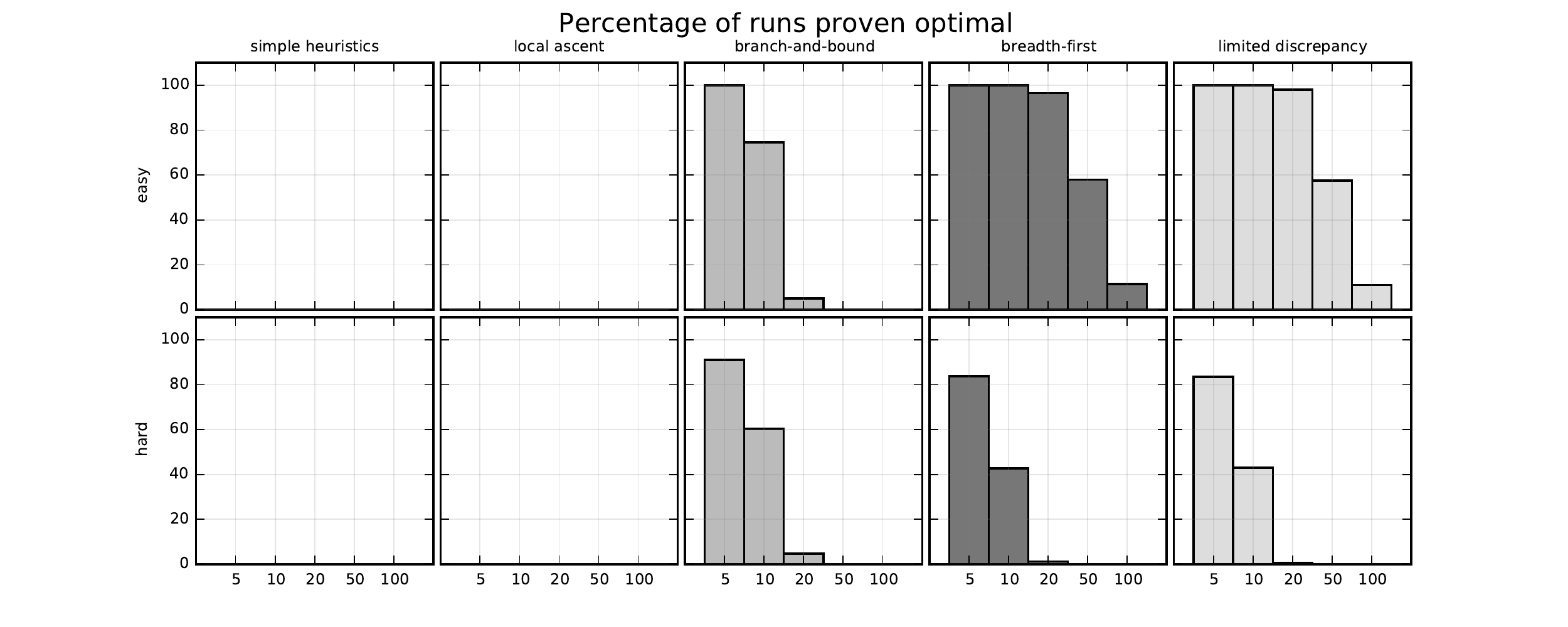}
    \includegraphics[width=.75\textwidth,trim=30 10 70 0,clip=True]{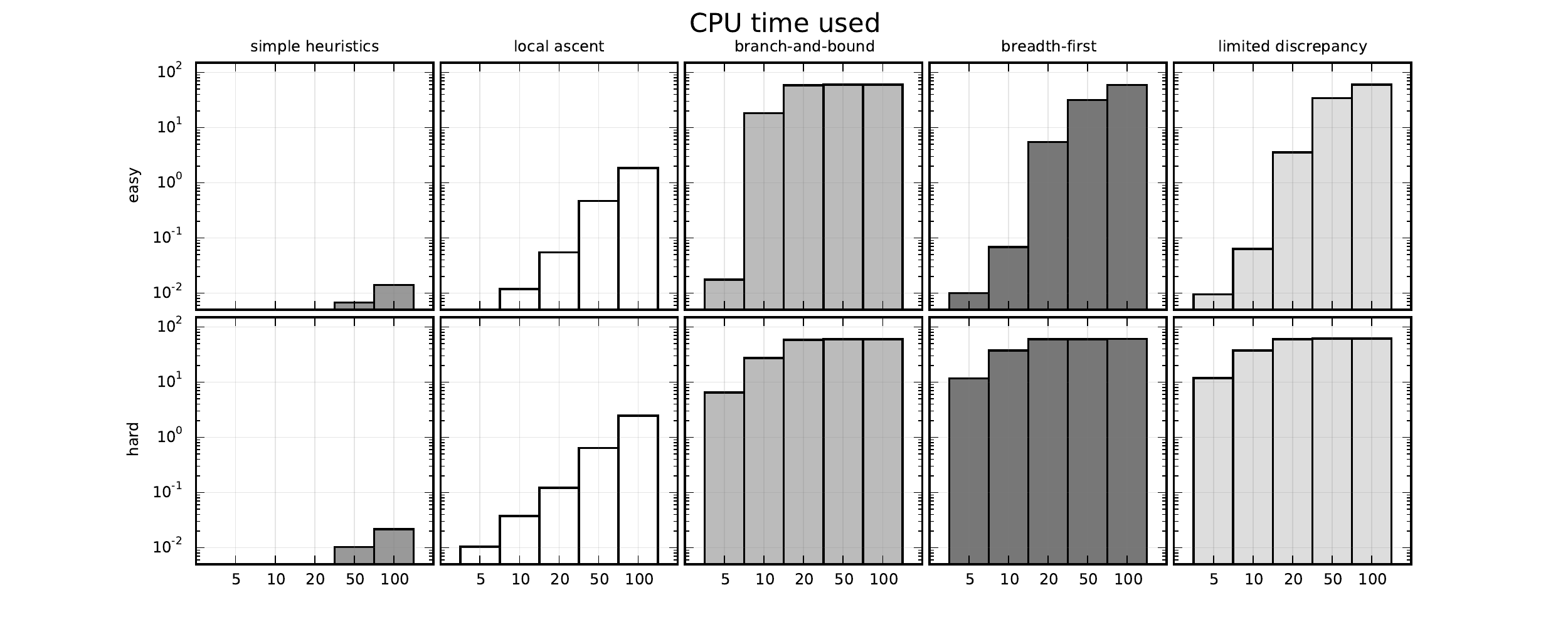}
    \includegraphics[width=.75\textwidth,trim=30 10 70 0,clip=True]{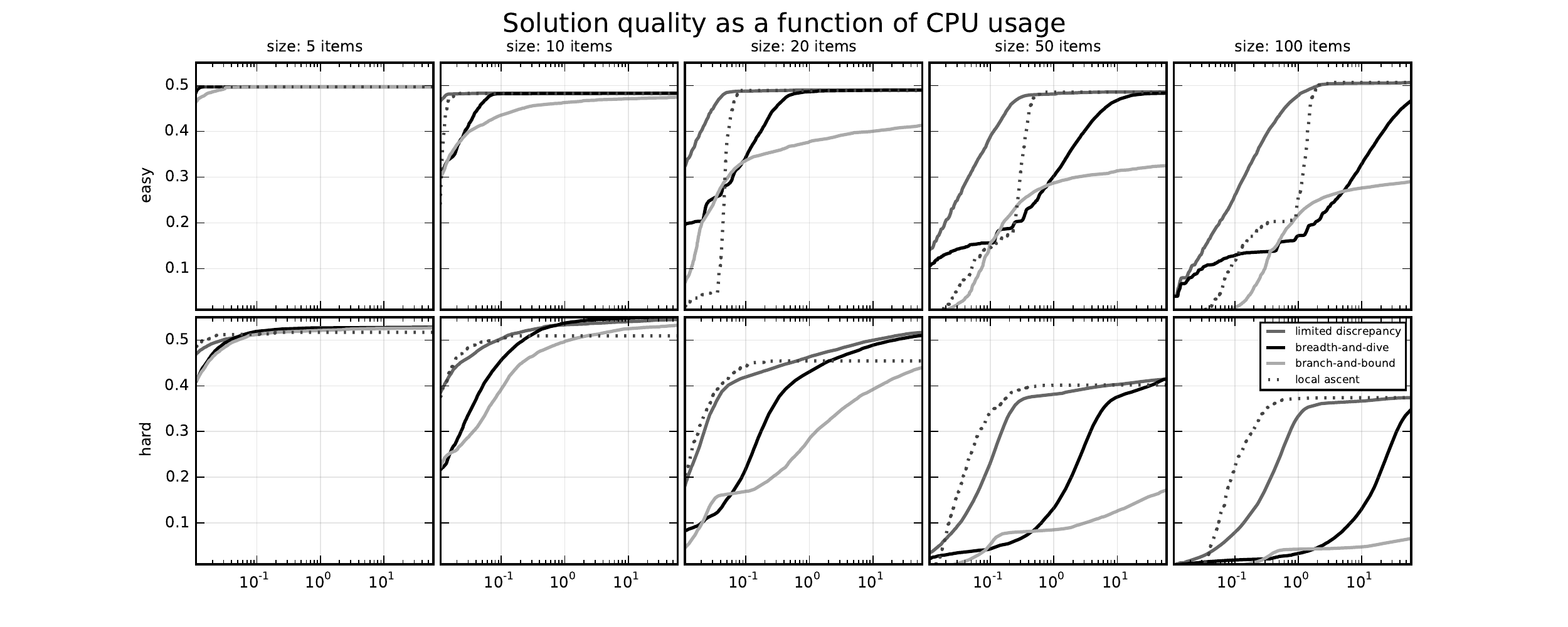}
  \end{center}
  \caption{Overall aper{\c c}u of the methods' performace: percentage of best solutions found (top), percentage of optimal executions (upper-center) and CPU time (lower-center) used, in terms of type (easy/hard) and number of items of the instance (measures are percentages/averages considering all instances of each size/type).  Bottom: evolution of the average of the objective value for all the instances of given size/type in terms of CPU time.}
  \label{fig:apercu}
\end{figure}

Figure~\ref{fig:apercu} graphically summarizes the results obtained.  On each sub-figure, results for instances of type easy and hard are separated into two rows.  Each bar (or curve, in the bottom sub-figure) represents percentages or averages considering all the instances of each size.  Methods considered are, as before, the simple heuristic rule (LBF), local ascent based on this rule, branch-and-bound, breadth-first search with diving, and limited discrepancy search; to each method corresponds a column in the top three sub-figures, and a line in the bottom sub-figure.  The abscissa for the three top sub-figures is the instance size, and for the bottom sub-figure is the CPU time used.  For each instance we have identified the best solution found by all the methods (which in some cases is optimal); on the top sub-figure, the ordinate is the percentage of instances for which each method finds such solution.  The next set of plots shows the percentage of instances for which the search tree was completely explored (for the relevant methods).  Follows a plot of the average CPU time used in the solution process, for all the instances of each size/type; the time for each run was limited to 60 seconds, but in many cases was smaller.   Finally, the sub-figure in the bottom shows the evolution of the average value of the objective for the best solution found by each method, in terms of the CPU time used; here we can observe how the gap between the different methods progresses.

As can be seen in Figure~\ref{fig:apercu}, ``standard'' branch-and-bound (taking the node with the highest upper bound at each iteration) quickly becomes very limited, when the instance size increases; this is due to the very high branching factor.  Crossing information on that figure with that of Table~\ref{tab:nnodes}, we see that when instance size increases, a very large number of nodes are open, but, due to the time limitation, only a small part of them can be explored.  This can be observed for all except smallest, easy instances.  For finding good solutions in a limited time, methods fully exploiting the heuristics (LA, BFS and LDS) have a much better performance.  Note that, for larger values of the CPU limit, the number of open nodes may have to be limited for avoiding memory overflow.

We have seen in Table~\ref{tab:all} that the method that is able to find strictly better solutions than the others for more instances is limited discrepancy search.  This is corroborated by the evolution over time of the average solution, for all instances of a given size, presented at the bottom of Figure~\ref{fig:apercu}.  The general tendency is to have LDS finding good solutions more quickly than the other tree search methods; however, near the CPU limit imposed, LDS is closely followed by  BFD (\eg, for hard instances of size 50).  In terms of the ability to complete the search, and hence to prove optimality, BFD and LDS are roughly equivalent; these methods appear to be considerably better than BB for easy instances, though slightly inferior for hard instances.

The main factor for LDS to be able to explore much more nodes than BFD is the ability to easily keep nodes organized by increasing discrepancy; for technical details, please consult the implementation code (see Appendix~\ref{sec:data}).      

Another interesting observation concerns the performance of local ascent.  For small instances, LA quickly finds the best solution (often proven optimal by tree search); however, LA is outclassed by tree search methods for mid-sized instances, to regain a relative good performance for large instances, as can be seen in the top graphic of Figure~\ref{fig:apercu}.  This is because local ascent is very fast, and hence the time constraint is not limiting it in our experiment, even for large instances.

\begin{table}
  \centering
  \caption{Average number of nodes explored, remaining in the queue at the end of the search, and created, for each of the tree search methods.}
  \label{tab:nnodes}       
\sisetup{
round-integer-to-decimal,
table-format = 5.0,
round-mode = places,
round-precision = 0,
}%
\begin{small}
\begin{tabular}{lr|*{3}S|*{3}S|*{3}S}
\multicolumn{2}{c|}{Instance}	& \multicolumn{3}{c|}{Nodes explored}	& \multicolumn{3}{c|}{Nodes in queue}	& \multicolumn{3}{c}{Nodes created} \\
Type	& Size	& {BB}	& {BFS}	& {LDS}	& {BB}	& {BFS}	& {LDS}	& {BB}	& {BFS}	& {LDS}\\ \hline
easy	& 5	& 10.38	& 5	& 25.96	& 0	& 0	& 0	& 27.84	& 16.545	& 16.695 \\
easy	& 10	& 3193.46	& 11.97	& 103.195	& 4379.28	& 0	& 0	& 13790	& 82.42	& 78.09 \\
easy	& 20	& 8113.42	& 165.94	& 1895.53	& 17430.8	& 998.705	& 3699.11	& 37510.8	& 3348.65	& 5502.44 \\
easy	& 50	& 1973.79	& 139.17	& 6052.78	& 17811.1	& 6167.86	& 109864	& 23525.5	& 8231.42	& 115697 \\
easy	& 100	& 574.07	& 54.485	& 4487.73	& 13399.5	& 7131.38	& 208455	& 14544.6	& 7811.54	& 212709 \\
hard	& 5	& 1895.21	& 4691.01	& 9091.66	& 1279.15	& 2830.75	& 4108.38	& 4563.49	& 9964.85	& 12813.2 \\
hard	& 10	& 5586.23	& 7454.29	& 28896.2	& 7979.26	& 12363.2	& 23270.6	& 20654& 32700.7	& 51551.2 \\
hard	& 20	& 4951	& 3429.5	& 34053.7	& 25272.2	& 22851	& 73126.3	& 40219.9	& 40261.3	& 106639 \\
hard	& 50	& 890.347	& 446.882	& 10487.9	& 26709.5	& 16993.2	& 268792	& 27603.2	& 18228.7	& 278513 \\
hard	& 100	& 290.43	& 83.8375	& 4649.9	& 18534.9	& 7564.85	& 256590	& 18824.5	& 7675.5	& 260851 \\
  \end{tabular}
\end{small}
\end{table}

\section{Conclusions}
\label{sec:conclusions}

Semifluids are materials having both fluid and solid characteristics.  In this paper, we studied the problem of packing a particular type of semifluid which cannot flow in one direction, though it is fluid in the other directions; this is the case when tubes of a small diameter are packed in parallel.  In this context, a packing item --- a semifluid --- is a set of identical tubes.  Different items have different length and/or value, and any fraction of an item may be used, with the objective of obtaining the maximum value packed.

Given the assumption of continuity, \ie, that one may arbitrarily divide a given volume, the problem of packing a set of tubes of different lengths in a container is surprisingly difficult.  This paper presents heuristics and complete search for the variant which closer corresponds to the industrial application: all the semifluids must be packed in the same direction, and a semifluid placed on top of another must not protrude.  In this paper, we have considered divisions of the volume of an item only when it reaches the ceiling of the container.

Several methods, from simple heuristics to complete search, are proposed.  The choice among them depends on the application. Simple heuristics are very quick, but often fail to find good solutions.  Local ascent based on simple heuristics often finds very good solutions, and is likely to be the best method for large instances and limited CPU time.  Among tree search methods, limited discrepancy search is often superior to the others, finding solutions of very good quality and frequently proving them optimal.

Semifluid packing under assumptions not considered in this paper is an interesting subject for future research; in particular, exploring different packing directions and the possibility of overflow.  The complexity class of this problem is unknown; determining it is an interesting research topic.  Another interesting research direction concerns developing compact mathematical models for optimization, taking full consideration of the possibility of packing fractions of each item.
The proposed heuristics could be extended and refined, in particular for taking into account the possibility of diversifying the point of division of an item into several packing places, further than the top of the container.  Yet another unexplored possibility for improvement concerns using the objective value of a heuristic solution as a lower bound, at the root node.

\subsection*{Acknowledgements}
This work is partly funded by FCT – Funda{\c c}{\~a}o para a Ci{\^e}ncia e a Tecnologia (Portuguese Foundation for Science and Technology) within project UID/EEA/50014/2013.
We would like to thank Benjamin M{\"u}ller, from Zuse Institute Berlin, for important suggestions.  We would also like to thank three anonymous reviewers for their constructive comments on a previous version of this paper.

\appendix

\section{Supplementary programs and data}
\label{sec:data}
Supplementary programs and data associated with this article can be found online at \url{http://www.dcc.fc.up.pt/~jpp/code/semifluid}.  That page contains an implementation of all the algorithms described in this paper and the program used for generating the instances, as well as the generated data.  

In the real-world application of this problem the number of different tube lengths in catalog is small.  To a smaller extent, this is also true for other numeric values in the required data.  We simulate this by limiting the number of digits in the random numbers generated: 3 digits in general, 2 or 3 digits for tube lengths.  All the values are normalized, so that the container dimensions are $1 \times 1 \times 1$.  As we use exact arithmetic for all operations, values generated and stored in the files are fractions.  The combinations of parameters used for instance generation are summarized in Table~\ref{tab:data}.

\begin{table}[h!tbp]
  \centering
  \caption{Characteristics of benchmark instances used: for each set of parameters 100 independent random instances have been generated, totaling 3000 instances.}
\begin{tabular}{lccccc}
  Type & Number of items    & Digits in $\ell_i$ & Volume of items (\% of $D \times W \times H $)  & Total \\ \hline
  easy & 5, 10, 20, 50, 100 & 2, 3               & 100\%                                           & 1000 instances\\
  hard & 5, 10, 20, 50, 100 & 2, 3               & 100\%, 150\%                                    & 2000 instances\\
\end{tabular}
  \label{tab:data}
\end{table}

For hard instances no optimum is known in advance.  These instances have been generated by simply drawing random numbers for lengths and volumes with the required number of digits, and afterwards updating the volumes so that the total volume will be the desired factor of the container's volume (in our data set, 1 or 3/2).  

Easy instances have the space of the container completely filled.  This is done by successive divisions of the container, as shown in Algorithm~\ref{alg:geneasy}.  To each holder generated this way there will correspond a different item.  Using this procedure, the total volume of items will always equal the volume of the container.

Instances closer to the real-world application that motivated this paper are hard instances with 10 to 20 items, two digits in $\ell_i$ and items occupying 100\% to 150\% of the container's volume.
\begin{algorithm}[htbp!]
    \DontPrintSemicolon
  create a holder $h$ with the size of the container\;
  $\setH \eq [h]$ \;
  \Repeat{number of holders is equal to the number of desired items}{
    randomly select a holder $h$ from $\setH$\;
    randomly select $r$ with uniform distribution in $[0,1]$\;
    \If(\tcp*[f]{With 50\% probability}){$r < 1/2$}{
      \If{$h$ has no other holders on top}{
        divide $h$ vertically\;
        replace $h$ by the two newly created holders
      }
    }
    \Else{
      divide $h$ horizontally\;
      replace $h$ by the two newly created holders
    }
  }
  \caption{Main steps for generating an easy instance.}\label{alg:geneasy}
\end{algorithm}

\bibliographystyle{spmpsci}      
\bibliography{occ.bib}   

\end{document}